\documentclass{article}
\usepackage{epsfig}
\title{Distinguishing two-qubit states using local measurements and
restricted classical communication}
\author{Mark Hillery and Jihane Mimih \\ Department of Physics and
Astronomy \\ Hunter College of CUNY \\ 695 Park Avenue \\ New York,
NY 10021}
\begin{document}
\maketitle
\begin{abstract}
The problem of unambiguous state discrimination consists of determining
which of a set of known quantum states a particular system is in.  One
is allowed to fail, but not to make a mistake.  The optimal procedure
is the one with the lowest failure probability.  This procedure has
been extended to bipartite states where the two parties, Alice and Bob,
are allowed to manipulate their particles locally and communicate
classically in order to determine which of two possible two-particle
states they have been given.  The failure probability of this local
procedure is the same as if the two particles were together in the same
location.  Here we examine the effect of restricting the classical
communication between the parties, either allowing none or eliminating
the possibility that one party's measurement depends on the result of
the other party's.  These issues are studied for two-qubit states, and
optimal procedures are found.  In some cases the restrictions cause
increases in the failure probability, but in other cases they do not.
Applications of these procedure, in particular to secret sharing, are
discussed.
\end{abstract}
\section{Introduction}
Suppose that we have a two-qubit state, and we give one of the qubits
to Alice and the other to Bob.  Alice and Bob know that the state
is either $|\Psi_{0}\rangle$ or $|\Psi_{1}\rangle$, and by making
local measurements and communicating classically, they want to
determine which state they have.  We want to consider the case of
unambiguous discrimination, which means that Alice and Bob may
fail to decide which state they have, but if they succeed, they
will not make an error.  That is, they will never conclude that
they have $|\Psi_{0}\rangle$ when they have been given
$|\Psi_{1}\rangle$ and vice versa.  Our object is to develop
a procedure that Alice and Bob can use to discriminate between
the states.

One aspect of this problem has already been solved.  If each state
is equally likely and and both qubits can be measured together,
then it is known that the states can be successfully unambiguously
discriminated with a probability of $p_{idp}=1-|\langle\Psi_{0}|
\Psi_{1}\rangle |$ \cite{ivanovic}-\cite{peres}.  It was recently
shown that the states can be discriminated using only local
operations and classical communication (LOCC) with the same
success probability.  Walgate, et al. proved that if
$\langle\Psi_{0}|\Psi_{1}\rangle =0$, then the states can be
distinguished perfectly  using only LOCC \cite{walgate}.
The case when $|\Psi_{0}\rangle$ and $|\Psi_{1}\rangle$ are not
orthogonal was investigated numerically by Virmani, et al.
\cite{virmani}, and they found strong evidence that unambiguous
discrimination is possible with a probability of $p_{idp}$ using
LOCC.  In addition, they found a class of states for which they
could prove that this was true.  A proof that this is true for
all bipartite states was provided by Chen and Yang \cite{chen}.

The procedure that makes LOCC unambiguous discrimination with
a success probability of $p_{idp}$ possible is the following.
Alice makes a projective measurement on her particle that
gives her no information about whether the state is
$|\Psi_{0}\rangle$ or $|\Psi_{1}\rangle$, and she then communicates
her result to Bob.  Based on what Alice has told him, Bob chooses
a measurement to make on his particle.  In particular, he applies
the procedure for the optimal unambiguous discrimination of
single qubit
states to his particle.  However, in this procedure one must know
the two states that one is discriminating between, and it is this
information that is provided by the result of Alice's measurement.

What we wish to examine here is how restricting the classical communication
between the parties affects their ability to discriminate between the
states.  We shall first see what happens when no classical communication
is allowed.  In that case each party has three possible measurement
results, $0$ corresponding to $|\Psi_{0}\rangle$, $1$ corresponding to
$|\Psi_{1}\rangle$, and $f$ for failure to distinguish.  If 
$|\Psi_{0}\rangle$ is sent, then Alice and Bob both measure $0$ or both
measure $f$, so that they both know, without communicating, that 
$|\Psi_{0}\rangle$ was sent or that the measurement failed.  If
$|\Psi_{1}\rangle$ is sent, then they both measure either $1$ or $f$.
We shall then relax the ban on classical commuincation, and allow Alice
and Bob to communicate their meaurement results to each other.  However,
conditional measurements will still be banned, i.e.\ situations in 
which the measurement made by one party depends on the measurement
results of the other will not be allowed.

One motivation for studying these situations, in addition to what
they tell us about state discrimination, is their possible use in 
communication schemes.
State discrimination for single qubits can be used to construct a
scheme for quantum cryptography \cite{bennett}.  In this protocol,
Alice and Bob wish to share a secure key.  Alice sends single
qubits to Bob in one of two nonorthogonal states, $|\psi_{0}\rangle$
or $|\psi_{1}\rangle$, and Bob applies the unambiguous state
discrimination protocol to the states he receives.  He then tells
Alice whether the procedure succeeded or failed, and they keep
the instances when it succeeded and throw out the rest.  If Bob's
measurement resulted in $|\psi_{0}\rangle$, then that particular
key bit is recorded as $0$, and if it resulted in
$|\psi_{1}\rangle$, it is recorded as $1$.  In this way a binary
string shared by Alice and Bob can be constructed, and it serves
as the key.  An eavesdropper, Eve, who intercepts the qubits that
Alice sends to Bob, and who wishes to find out which state they
are in, has a problem.  Because the states are not orthogonal,
she will not be able to definitely determine the state each of
each qubit she receives.  However, she must send a qubit in
either $|\psi_{0}\rangle$ or $|\psi_{1}\rangle$ on to Bob.  Since
her information about the qubit she received is not perfect,
Eve will sometimes send a qubit in the wrong state to Bob.  If
Alice and Bob publicly compare some of their key bits and find
discrepencies, then they know an eavesdropper was present.  If
they find no discrepencies, then they can conclude that they
share a secure key.

The no-classical-communication scheme would allow a third party, Charlie,
to distribute a shared key to Alice and Bob.  Charlie would send one 
qubit to Alice and one to Bob, where the qubits are either in the
state $|\Psi_{0}\rangle$ or $|\Psi_{1}\rangle$, and Alice and Bob
would measure them.  They would both know when the had found 
$|\Psi_{0}\rangle$, when they had found $|\Psi_{1}\rangle$, and
when they had failed.  Note that Charlie would not know the key,
because he would not know which bits corresponded to failure.  A
slight relaxation of the no-classical-communication condition allows
all three parties to share a key.  Alice and Bob simply announce
publicly when they failed to distinguish the state.

A possible use for the second scheme, when Alice and Bob are allowed
to compare their meaurement results, is secret sharing.
In this case a third party, Charlie, wants to share a secure
key with Alice and Bob, but he wants Alice and Bob to have to
cooperate to determine the key bit.  Neither Alice nor Bob,
separately, will know the key, but together they will.  Charlie
accomplishes this by sending one qubit to Alice and another to
Bob.  The two qubits are either in the state $|\Psi_{0}\rangle$
or $|\Psi_{1}\rangle$, and these states are not orthogonal.
Alice and Bob then perform a procedure to determine which state
they have, and this procedure must require their cooperation,
so that neither of them by themselves can determine the state.
If they use the optimal procedure in which the measurement Bob
makes depends on the result of Alice's measurement, then Alice would 
measure her particle, and Bob would do nothing to his.  When they
want to determine the key bit, Alice will tell Bob the result of
her measurement, and Bob will make the appropriate measurement
on his particle.  This method, however, requires Bob to store
quantum information, i.e. keep his particle free from the effects
of decoherence, until the bit is determined.  A more practical
procedure would be the restricted-classical-communication scheme
in which both Alice and Bob make independent measurements,
and are able to determine the state from the results.  In that case, they
each measure their qubit when they receive it, and they record the 
results of their measurements.  This means that it is only classical
information that needs to be stored.  Neither Alice nor Bob should be
able to determine the state from just their own result, but by
putting their results together they should be able to indentify the state
they were sent with some nonzero probability, and they should never make 
an error.  It is this kind of procedure we wish to study here.

\section{No classical communication}
As discussed in the Introduction, we shall assume that Alice and Bob
each has one of three measurement alternatives, $0$, $1$, and $f$.  The
POVM operators that characterize the measurements are $\{ A_{0},A_{1},
A_{f}\}$ for Alice and $\{ B_{0},B_{1},B_{f}\}$ for Bob.  These operators
satisfy
\begin{equation}
\label{ident1}
I_{A}= \sum_{j=0,1,f}A_{j}^{\dagger}A_{j}\hspace{1cm} I_{B}=\sum_{j=0,1,f}
B_{j}^{\dagger}B_{j} ,
\end{equation}
where $I_{A}$ is the identity on $\mathcal{H}_{A}$, the Hilbert space of
Alice's qubit, and $I_{B}$ is the identity on $\mathcal{H}_{B}$, the 
space of Bob's qubit.  The requirement that Alice and Bob only get the
same result for their measurements imposes the conditions
\begin{equation}
A_{j}B_{k}|\Psi_{n}\rangle =0 ,
\end{equation}
where $j,k\in \{ 0,1,f\}$ and $j\neq k$, and $n\in \{ 0,1\}$.  In addition,
the fact that no errors are made in identitfying the states requires
that
\begin{equation}
A_{0}B_{0}|\Psi_{1}\rangle \hspace{1cm} A_{1}B_{1}|\Psi_{0}\rangle =0 .
\end{equation}

It is clear simply from the number of conditions, that if this procedure is
possible at all, it will be true only for a very restricted set of states.
In fact, what we find is that the best we can do is to is to identify one
of the states with a nonzero probability and fail the rest of the time.
The details of the proof of this statement are given in the Appendix.

We conclude this section with an example of the situation in which one
state can be detected.  Suppose our two states are given by
\begin{eqnarray}
|\Psi_{0}\rangle & = & |0\rangle |0\rangle \nonumber \\
|\Psi_{1}\rangle & = & \frac{1}{\sqrt{2}}(|0\rangle |0\rangle
+|1\rangle |1\rangle ) ,
\end{eqnarray}
where Alice's states are first and Bob's second.  In addition, we have 
that $A_{0}=B_{0}=0$, so that $|\Psi_{0}\rangle$ is never detected, and
\begin{eqnarray}
A_{1} = |1\rangle\langle 1| & B_{1}=|1\rangle\langle 1| \nonumber \\
A_{f} = |0\rangle\langle 0| & B_{f}=|0\rangle\langle 0| .
\end{eqnarray}
From this we see that, indeed, if $|\Psi_{0}\rangle$ is sent, then it
will not be detected, but if $|\Psi_{1}\rangle$ is sent, then we will
detect it with a probability of $1/2$ and fail with a probability of
$1/2$.  Thus, we are very limited in distinguishing two states without
any classical communication between Alice and Bob.

\section{Limited classical communication}
The situation becomes more interesting if we allow Alice and Bob
to communicate the results of their measurements to each other
only after both measurements have been made.
We now consider the following situation.  Alice and Bob make
measurements on their particles, and each of these measurements
can have one of two outcomes, $0$ or $1$.  Alice's measurement
is described by the POVM $\{ A_{0}, A_{1}\}$ and Bob's by
$\{ B_{0}, B_{1}\}$, where
\begin{equation}
\label{id}
I_{A}=A_{0}^{\dagger}A_{0}+A_{1}^{\dagger}A_{1}\hspace{1cm}
I_{B}=B_{0}^{\dagger}B_{0}+B_{1}^{\dagger}B_{1}  ,
\end{equation}
and $I_{A}$ and $I_{B}$ are the identity operators in Alice's
and Bob's Hilbert spaces, respectively.  The probability that
Alice will obtain the result $k$ if the two qubit-state is
$|\Psi_{j}\rangle$ is ${\rm Tr}(\rho_{Aj}A^{\dagger}_{k}A_{k})$,
where $\rho_{Aj}={\rm Tr}_{B}(|\Psi_{j}\rangle\langle\Psi_{j}|)$
is the reduced density matrix of $|\Psi_{j}\rangle$ in Alice's
space.  Similar expressions hold for the probabilities of Bob's
measurements.  Note that $A_{0}$ and $A_{1}$ commute with $B_{0}$
and $B_{1}$.

Together, Alice and Bob have four possible sets of results
(Alice's result is given first, Bob's second),
$\{ 0,0\}$, $\{ 0,1\}$, $\{ 1,0\}$, $\{ 1,1\}$, and we have to
decide which sets correspond to $|\Psi_{0}\rangle$, which to
$|\Psi_{1}\rangle$, and which to failure to decide.  Let us
first consider what happens if we assume that none of the
sets corresponds to failure.  In particular, suppose that
$\{ 0,0\}$ and $\{ 1,1\}$ correspond to $|\Psi_{0}\rangle$
and $\{ 0,1\}$ and $\{ 1,0\}$ correspond to $|\Psi_{1}\rangle$.
This implies that if the state is $|\Psi_{1}\rangle$, then
the probability of getting $\{ 0,0\}$ or $\{ 1,1\}$ is zero,
and if the state is $|\Psi_{0}\rangle$, the probability of
getting $\{ 0,1\}$ or $\{ 1,0\}$ is zero.  Therefore, we have
\begin{eqnarray}
\langle\Psi_{0}|A_{0}^{\dagger}A_{0}B_{1}^{\dagger}B_{1}|\Psi_{0}
\rangle & = \langle\Psi_{0}|A_{1}^{\dagger}A_{1}B_{0}^{\dagger}
B_{0}|\Psi_{0}\rangle & = 0 \nonumber \\
\langle\Psi_{1}|A_{0}^{\dagger}A_{0}B_{0}^{\dagger}B_{0}|\Psi_{1}
\rangle & = \langle\Psi_{1}|A_{1}^{\dagger}A_{1}B_{1}^{\dagger}
B_{1}|\Psi_{1}\rangle & = 0 .
\end{eqnarray}
These imply the simpler equations
\begin{eqnarray}
A_{0}B_{1}|\Psi_{0}\rangle & = A_{1}B_{0}|\Psi_{0}\rangle &
=0  \nonumber \\
A_{0}B_{0}|\Psi_{1}\rangle & = A_{1}B_{1}|\Psi_{1}\rangle &
=0 .
\end{eqnarray}
If we now note that
\begin{eqnarray}
\label{overlap}
\langle\Psi_{1}|\Psi_{0}\rangle & = & \langle\Psi_{1}|
I_{A}\otimes I_{B}|\Psi_{0}\rangle \nonumber \\
 & = & \langle\Psi_{1}|(A_{0}^{\dagger}A_{0}+A_{1}^{\dagger}
A_{1})\otimes (B_{0}^{\dagger}B_{0}+B_{1}^{\dagger}B_{1})
\Psi_{0}\rangle ,
\end{eqnarray}
we see from the previous equation that $\langle\Psi_{1}|
\Psi_{0}\rangle =0$.  Therefore, if we are able to distinguish
the states every time without error, they must be orthogonal.

Now let us suppose that some of the measurement results correspond
to a failure to distinguish the states.  We will focus on two
different cases.  In the first we shall assume that two of the four
alternatives correspond to failure, and in the second we shall assume
that only one does.

\subsection{Two failure states}
Let us assume that $\{ 0,0\}$ corresponds to
$|\Psi_{0}\rangle$, $\{ 1,1\}$ corresponds to $|\Psi_{1}\rangle$,
and both $\{ 0,1\}$ and $\{ 1,0\}$ correspond to failure to
distinguish. The condition of no errors implies that
\begin{equation}
\label{cond}
A_{0}B_{0}|\Psi_{1}\rangle = 0 \hspace{1cm}
A_{1}B_{1}|\Psi_{1}\rangle = 0 .
\end{equation}
If we apply these conditions to Eq. (\ref{overlap}), we find that
\begin{equation}
\label{conscond}
\langle\Psi_{1}|\Psi_{0}\rangle = \langle\Psi_{1}|F|\Psi_{0}\rangle ,
\end{equation}
where
\begin{equation}
\label{defF}
F=A_{0}^{\dagger}A_{0}B^{\dagger}_{1}B_{1}+A_{1}^{\dagger}
A_{1}B^{\dagger}_{0}B_{0}.
\end{equation}

Now let us examine the conditions in Eq. (\ref{cond}) in more detail.
We first express $|\Psi_{1}\rangle$ in its Schmidt basis
\begin{equation}
|\Psi_{1}\rangle = \sum_{j=0}^{1}\sqrt{\lambda_{1j}}|v_{Aj}\rangle
\otimes |v_{Bj}\rangle ,
\end{equation}
where $\{ v_{A0}, v_{A1}\}$ and $\{ v_{B0}, v_{B1}\}$ are orthonormal
bases for Alice's and Bob's spaces, respectively, and $\lambda_{1j}$
for $j=0,1$ are the eigenvalues of the reduced density matrixes.
The condition $A_{0}B_{0}|\Psi_{1}\rangle = 0$ then implies that
\begin{equation}
\sqrt{\lambda_{10}}A_{0}|v_{A0}\rangle\otimes B_{0}|v_{B0}\rangle =
-\sqrt{\lambda_{11}}A_{0}|v_{A1}\rangle\otimes B_{0}|v_{B1}\rangle .
\end{equation}
The only way this can be true is if $A_{0}|v_{A0}\rangle$ is parallel
to $A_{0}|v_{A1}\rangle$ and $B_{0}|v_{B0}\rangle$ is parallel to
$B_{0}|v_{B1}\rangle$.  Therefore, we can write
\begin{eqnarray}
\label{con1}
A_{0}|v_{A0}\rangle = c_{0}|\eta_{A}\rangle & B_{0}|v_{B0}\rangle
= d_{0}|\eta_{B}\rangle \nonumber \\
A_{0}|v_{A1}\rangle = c_{1}|\eta_{A}\rangle & B_{0}|v_{B1}\rangle
= d_{1}|\eta_{B}\rangle ,
\end{eqnarray}
where $c_{j}$ and $d_{j}$ are constants and $\|\eta_{A}\| =
\|\eta_{B}\| =1$.  These equation imply that
\begin{eqnarray}
A_{0}=\sum_{j=0}^{1}c_{j}|\eta_{A}\rangle\langle v_{Aj}|=
|\eta_{A}\rangle\langle r_{A}| \nonumber \\
B_{0}=\sum_{j=0}^{1}d_{j}|\eta_{B}\rangle\langle v_{Bj}|=
|\eta_{B}\rangle\langle r_{B}| ,
\end{eqnarray}
where
\begin{equation}
|r_{A}\rangle = \sum_{j=0}^{1}c^{\ast}_{j}|v_{Aj}\rangle
\hspace{1cm}
|r_{B}\rangle = \sum_{j=0}^{1}d^{\ast}_{j}|v_{Bj}\rangle .
\end{equation}
The condition $A_{0}B_{0}|\Psi_{1}\rangle = 0$ can now be
expressed as
\begin{equation}
\label{condit1}
(\langle r_{A}|\otimes \langle r_{B}|)|\Psi_{1}\rangle = 0.
\end{equation}

We can now do the same thing with the condition that $A_{1}B_{1}
|\Psi_{0}\rangle =0$.  Expressing $|\Psi_{0}\rangle$ in its
Schmidt basis we have that
\begin{equation}
|\Psi_{0}\rangle = \sum_{j=0}^{1}\sqrt{\lambda_{0j}}|u_{Aj}\rangle
\otimes |u_{Bj}\rangle ,
\end{equation}
where $\{ u_{A0}, u_{A1}\}$ and $\{ u_{B0}, u_{B1}\}$ are orthonormal
bases for Alice's and Bob's spaces, respectively, and $\lambda_{0j}$
for $j=0,1$ are the eigenvalues of the reduced density matrixes.
Applying the same reasoning as before, we find that
\begin{equation}
A_{1}=|\xi_{A}\rangle\langle s_{A}| \hspace{1cm} B_{1}=
|\xi_{B}\rangle\langle s_{B}| ,
\end{equation}
where $\|\xi_{A}\| = \|\xi_{B}\| = 1$.  We also have that
\begin{equation}
\label{condit2}
(\langle s_{A}|\otimes \langle s_{B}|)|\Psi_{0}\rangle = 0.
\end{equation}

We can gain more information about the vectors $|r_{A}\rangle$,
$|r_{B}\rangle$, $|s_{A}\rangle$, and $|s_{B}\rangle$ by
substituting the results of the previous paragraphs into Eqs.
(\ref{id}).  This gives us that
\begin{equation}
\label{id2}
I_{A}=|r_{A}\rangle\langle r_{A}|+|s_{A}\rangle\langle s_{A}|
\hspace{1cm} I_{B}=|r_{B}\rangle\langle r_{B}|+|s_{B}\rangle
\langle s_{B}| .
\end{equation}
Now let both sides of the first of these equations act on the
vector $|r_{A}\rangle$,
\begin{equation}
\| r_{A}\|^{2}|r_{A}\rangle + |s_{A}\rangle\langle s_{A}|r_{A}
\rangle = |r_{A}\rangle .
\end{equation}
The only way this can be true is if either $|r_{A}\rangle$ is
parallel to $|s_{A}\rangle$ which violates Eq. (\ref{id2}),
or if $\langle s_{A}|r_{A}\rangle = 0$ and $\| r_{A}\| =1$.
Therefore, $|s_{A}\rangle$ is orthogonal to $|r_{A}\rangle$,
and both have norm $1$.  Henceforth, we shall denote $|s_{A}
\rangle$ by $|r_{A}^{\perp}\rangle$, and we have that
$\{ r_{A}, r_{A}^{\perp}\}$ is an orthonormal basis for
Alice's space.  Similarly, we find that $\{ r_{B},
r_{B}^{\perp}\}$, where $|r_{B}^{\perp}\rangle =|s_{B}\rangle$,
is an orthonormal basis for Bob's space.

Now let us examine the failure probabilities.  We first express
the operator $F$, defined in Eq. (\ref{defF}) as
\begin{eqnarray}
F &=& (|r_{A}\rangle\otimes |r_{B}^{\perp}\rangle )(\langle r_{A}|
\otimes \langle r_{B}^{\perp}| )+(|r_{A}^{\perp}\rangle\otimes
|r_{B}\rangle )(\langle r_{A}^{\perp}| \otimes \langle r_{B}|)
\nonumber \\
 & = & I-(|r_{A}\rangle\otimes |r_{B}\rangle )(\langle r_{A}|
\otimes \langle r_{B}| ) \nonumber \\
 & & - (|r_{A}^{\perp}\rangle\otimes |r_{B}^{\perp}\rangle )
(\langle r_{A}^{\perp}|\otimes \langle r_{B}^{\perp}| ) .
\end{eqnarray}
We first note that if Eqs. (\ref{condit1}) and (\ref{condit2}) are
satisfied, then the condition in Eq. (\ref{conscond}) is also
satisfied.  The failure probability if Charlie sends the state
$|\Psi_{0}\rangle$ is $\langle\Psi_{0}|F|\Psi_{0}\rangle$, and
if he sends the state $|\Psi_{1}\rangle$, it is
$\langle\Psi_{1}|F|\Psi_{1}\rangle$.  These probabilities can
be expressed as
\begin{eqnarray}
\label{fail1}
\langle\Psi_{0}|F|\Psi_{0}\rangle & = & 1-|(\langle r_{A}|
\otimes \langle r_{B}| )|\Psi_{0}\rangle |^{2} \nonumber \\
\langle\Psi_{1}|F|\Psi_{1}\rangle & = & 1-|(\langle r_{A}^{\perp}|
\otimes \langle r_{B}^{\perp}| )|\Psi_{1}\rangle |^{2} .
\end{eqnarray}
If each of the states is equally likely, then the total failure probability,
$p_{f}$, is given by
\begin{equation}
\label{fail2}
p_{f}=\frac{1}{2}(\langle\Psi_{0}|F|\Psi_{0}\rangle +
\langle\Psi_{1}|F|\Psi_{1}\rangle ) .
\end{equation}
We want to minimize this overall failure probability.

Note that the failure probabilities are unaffected by the choice of the
vectors $|\xi_{A}\rangle , |\xi_{B}\rangle , |\eta_{A}\rangle$, and
$|\eta_{B}\rangle$.  If we make the choices
\begin{eqnarray}
|\xi_{A}\rangle = |r_{A}^{\perp}\rangle & |\xi_{B}\rangle = |r_{B}^{\perp}
\rangle \nonumber \\
|\eta_{A}\rangle = |r_{A}\rangle & |\eta_{B}\rangle = |r_{B}\rangle ,
\end{eqnarray}
then the operators $A_{j}$ and $B_{j}$, where $j=1,2$, are projections
and the generalized meausrement becomes a von Neumann measurement.

Let us summarize our remaining problem.  We want to find a basis
for Alice's space, $\{ |r_{A}\rangle , |r_{A}^{\perp}\rangle\}$,
and one for Bob's space, $\{ |r_{B}\rangle , |r_{B}^{\perp}
\rangle\}$, that satisfy the conditions
\begin{eqnarray}
(\langle r_{A}^{\perp}|\otimes \langle r_{B}^{\perp}|)
|\Psi_{0}\rangle  & = & 0 \nonumber \\
(\langle r_{A}|\otimes \langle r_{B}|)|\Psi_{1}\rangle & = & 0 .
\end{eqnarray}

We can reduce these conditions to the solution of several simple
equations.  First, expanding $|r_{A}^{\perp}\rangle$ and 
$|r_{B}^{\perp}\rangle$ in terms of $|u_{Aj}\rangle$ and
$|u_{Bj}\rangle$, respectively, we have
\begin{equation}
|r_{A}^{\perp}\rangle = \sum_{j=0}^{1}e^{\ast}_{j}|u_{Aj}\rangle
\hspace{1cm}
|r_{B}^{\perp}\rangle = \sum_{j=0}^{1}f^{\ast}_{j}|u_{Bj}\rangle .
\end{equation}
The equations in the previous paragraph become
\begin{equation}
\label{statcon}
\sum_{j=0}^{1}\sqrt{\lambda_{0j}}e_{j}f_{j}=0 \hspace{1cm}
\sum_{j=0}^{1}\sqrt{\lambda_{1j}}c_{j}d_{j}=0 ,
\end{equation}
while the conditions $\langle r_{A}^{\perp}|r_{A}\rangle =0$
and $\langle r_{B}^{\perp}|r_{B}\rangle =0$ become
\begin{eqnarray}
\label{perpcon}
\sum_{j_{1},j_{2}=0}^{1}c_{j_{1}}e_{j_{2}}^{\ast}\langle
v_{Aj_{1}}|u_{Aj_{2}}\rangle =0 \nonumber \\
\sum_{j_{1},j_{2}=0}^{1}d_{j_{1}}f_{j_{2}}^{\ast}\langle
v_{Bj_{1}}|u_{Bj_{2}}\rangle =0 .
\end{eqnarray}
Now define the ratios
\begin{eqnarray}
z_{1}=\frac{c_{1}^{\ast}}{c_{0}^{\ast}} & z_{2}=\frac{d_{1}^{\ast}}
{d_{0}^{\ast}} \nonumber \\
z_{3}=\frac{e_{1}^{\ast}}{e_{0}^{\ast}} & z_{1}=\frac{f_{1}^{\ast}}
{f_{0}^{\ast}} .
\end{eqnarray}
If we now divide Eqs.\ (\ref{statcon}) and (\ref{perpcon}) by the
appropriate product of expansion coefficients, e.g.\ the first of
Eqs.\ (\ref{statcon}) is divided by $e_{0}f_{0}$ and the first of
Eqs.\ (\ref{perpcon}) is divided by $c_{0}e_{0}^{\ast}$, we find
\begin{eqnarray}
\label{conditions1}
\sqrt{\lambda_{00}}+\sqrt{\lambda_{01}}z_{3}z_{4}=0 \nonumber \\
\sqrt{\lambda_{10}}+\sqrt{\lambda_{11}}z_{1}z_{2}=0 \nonumber \\
\langle v_{A0}|u_{A0}\rangle +\langle v_{A0}|u_{A1}\rangle z_{3} 
\nonumber \\
+\langle v_{A1}|u_{A0}\rangle z_{1}^{\ast}+\langle v_{A1}|u_{A1}\rangle
z_{1}^{\ast}z_{3} =0 \nonumber \\
\langle v_{B0}|u_{B0}\rangle +\langle v_{B0}|u_{B1}\rangle z_{4} 
\nonumber \\
+\langle v_{B1}|u_{B0}\rangle z_{2}^{\ast}+\langle v_{B1}|u_{B1}\rangle
z_{2}^{\ast}z_{4} =0 .
\end{eqnarray}
Given two specific states $|\Psi_{0}\rangle$ and $|\Psi_{1}\rangle$, these
equations can be solved to find the vectors $|r_{A}\rangle$, $|r_{A}^{\perp}
\rangle$, $|r_{B}\rangle$, and $|r_{B}^{\perp}\rangle$.

Let us now consider two examples. In the first we shall suppose 
that $|\Psi_{0}\rangle$ and $|\Psi_{1}\rangle$ have the same
Schmidt bases while in the second the Schmidt bases of the two 
states will be different.

We begin by assuming that our two states are given by
\begin{eqnarray}
\label{samebasis}
|\Psi_{0}\rangle =  \cos\theta_{0}|00\rangle +
\sin\theta_{0}|11\rangle \nonumber \\ 
|\Psi_{1}\rangle = \cos\theta_{1}|00\rangle + \sin\theta_{1}
|11\rangle ,
\end{eqnarray}
where $\theta_{0}$ and $\theta_{1}$ are both between $0$ and $\pi /2$.
Solving Eqs.\ (\ref{conditions1}) for these states, we first find the 
condition $\tan\theta_{0}\tan\theta_{1}=1$, which implies that $\theta_{1}
=(\pi /2)-\theta_{0}$.  We also find explicit expressions for the vectors
\begin{eqnarray}
|r_{A}\rangle = c^{\ast}_{0}  ( |0\rangle + z_{1} |1\rangle )\nonumber \\
|r_{B}\rangle = d^{\ast}_{0}  \left(  |0\rangle
-\frac{\cot\theta_{1}}{z_{1}}|1\rangle\right)\nonumber \\
|r_{A}^{\perp}\rangle = e^{\ast}_{0}\left( |0\rangle -\frac{1}{z_{1}}
|1\rangle\right)\nonumber \\
|r_{B}^{\perp}\rangle = f^{\ast}_{0}  ( |0\rangle +
\cot\theta_{0} z^{\ast}_{1} |1\rangle ) ,
\end{eqnarray}
where the normalization constants are given by
\begin{eqnarray}
|c_{0}|^{2}=\frac{1}{1+|z_{1}|^{2}}\nonumber \\
|d_{0}|^{2}=\frac{|z_{1}|^{2}}{|z_{1}|^{2}+(\cot\theta_{1})^{2}}\nonumber \\
|e_{0}|^{2}=\frac{|z_{1}|^{2}}{1+|z_{1}|^{2}}\nonumber \\
|f_{0}|^{2}=\frac{1}{1+(\cot\theta_{1})^{2}|z_{1}|^{2}} .
\end{eqnarray}
The quantity $z_{1}$ is at the moment undetermined, but it will be fixed by
requiring the failure probability to be a minimum.  This probability is now
given by
\begin{equation}
\label{fail3}
p_{f}=1-\frac{|z_{1}|^{2}}{2+2|z_{1}|^{2}}\
\frac{(1-(\cot\theta_{1})^{2})^{2}}{1+(|z_{1}|\cot\theta_{1})^{2}}\ 
\frac{(1-(\tan\theta_{1})^{2})^{2}}{1+(|z_{1}|\tan\theta_{1})^{2}} ,
\end{equation}
where the condition $\theta_{0}=(\pi /2)-\theta_{1}$ has been used to
eliminate $\theta_{0}$.  Setting the derivative of $p_{f}$ with respect
to $|z_{1}|^{2}$ equal to zero, we find an equation that has only one
positive solution, $|z_{1}|^{2}=\cot\theta_{1}$.  Substituting this
value into Eq.\ (\ref{fail3}), we find
\begin{equation}
p_{f}=\sin (2\theta_{1})
\end{equation}

This failure probability should be compared to that when a single joint 
measurement can be performed on both qubits of the two-qubit states.
In that case, if each of the states is equally likely, then the probability
of failing to distinguishing the states is given by the IDP limit
\begin{equation}
p_{fidp}= |\langle\Psi_{0}|\Psi_{1}\rangle |=\sin (2\theta_{1}) .
\end{equation}
Note that this expression is identical to that given in the previous
paragraph.  Therefore, in this example we can conclude that the failure
probability that is achieved by measuring the qubits separately is the
same as that when the qubits are measured together.

Now let us see what happens if the states have different Schmidt bases.
We shall keep $|\Psi_{0}\rangle$ as before, but choose $|\Psi_{1}\rangle$
differently, 
\begin{eqnarray}
|\Psi_{0}\rangle = \cos\theta_{0}|00\rangle +\sin\theta_{0}|11\rangle
\nonumber \\
|\Psi_{1}\rangle = \cos\theta_{1}|+x\rangle |+x\rangle +
\sin\theta_{1}|-x\rangle |-x\rangle ,
\end{eqnarray}
where $|\pm x\rangle= (1/\sqrt{2})(|0\rangle \pm |1\rangle)$.  Solving
Eqs.\ (\ref{conditions1}) for these states, we first find a quadratic
equation for $z_{1}$
\begin{equation}
(1-\cot\theta_{0})z_{1}^{2}-(1-\cot\theta_{1})(1+\cot\theta_{0})z_{1}
-(1-\cot\theta_{0})\cot\theta_{1}=0 .
\end{equation}
The vectors making up the POVM are given by
\begin{eqnarray}
|r_{A}\rangle &=& c_{0}^{\ast}(|+x\rangle +z_{1}|-x\rangle ) \nonumber \\
|r_{B}\rangle &=& d_{0}^{\ast}\left( |+x\rangle -\frac{\cot\theta_{1}}
{z_{1}}|-x\rangle \right)  \nonumber \\
|r_{A}^{\perp}\rangle &=& e_{0}^{\ast}(|0\rangle +z_{3}|1\rangle )
\nonumber \\
|r_{B}\rangle &=& f_{0}^{\ast}\left( |0\rangle -\frac{\cot\theta_{0}}
{z_{3}}|1\rangle \right) .
\end{eqnarray}
The normalization constants are given by
\begin{eqnarray}
|c_{0}|^{2}=\frac{1}{1+|z_{1}|^{2}} & |e_{0}|^{2}=\frac{1}{1+|z_{3}|^{2}}
\nonumber \\
|d_{0}|^{2}=\frac{|z_{1}|^{2}}{|z_{1}|^{2}+\cot^{2}\theta_{1}} &
|f_{0}|^{2}=\frac{|z_{3}|^{2}}{|z_{3}|^{2}+\cot^{2}\theta_{0}},
\end{eqnarray}
where
\begin{equation}
z_{3}= -\frac{1-\cot\theta_{0}\cot\theta_{1}+(1-\cot\theta_{0})z_{1}^{\ast}}
{1-\cot\theta_{1}} .
\end{equation}
The failure probability is given by Eqs.\ (\ref{fail1}) and (\ref{fail2}),
where
\begin{eqnarray}
|(\langle r_{A}|\otimes\langle r_{B}|)\Psi_{0}\rangle |^{2}=\frac
{|z_{1}|^{2}\sin^{2}\theta_{0}}{4(1+|z_{1}|^{2})(|z_{1}|^{2}+\cot^{2}
\theta_{1})} \nonumber \\
\left| (1+\cot\theta_{0})(1-\cot\theta_{1})+(\cot\theta_{0}-1)\left( 
z_{1}^{\ast}-\frac{\cot\theta_{1}}{z_{1}^{\ast}}\right)\right|^{2} 
\nonumber \\
|(\langle r_{A}^{\perp}|\otimes\langle r_{B}^{\perp}|)\Psi_{0}\rangle |^{2}
=\frac{|z_{3}|^{2}\sin^{2}\theta_{1}}{4(1+|z_{3}|^{2})(|z_{1}|^{3}+\cot^{2}
\theta_{0})} \nonumber \\
\left| (1+\cot\theta_{1})(1-\cot\theta_{0})+(\cot\theta_{1}-1)\left( 
z_{3}-\frac{\cot\theta_{0}}{z_{3}}\right)\right|^{2} .
\end{eqnarray}

Specializing to the case $\theta_{0}=\pi /2$ we find that there are two
sets of values for $z_{1},\ldots z_{4}$.  One set is obtained from the
other simply by reversing the roles of $|r_{A}\rangle$ and $|r_{B}\rangle$,
and both give the same failure probability, so that we need only consider
one of them.  Doing so we have that
\begin{eqnarray}
z_{1}=\cot\theta_{1}& \hspace{1cm}& z_{2}=-1 \nonumber \\
z_{3}=\frac{1+\cot\theta_{1}}{\cot\theta_{1}-1} &\hspace{1cm} & z_{4}=0 . 
\end{eqnarray}
This gives a value for the failure probability of
\begin{equation}
p_{f}=1-\frac{(1-\cot\theta_{1})^{2}+(\cos\theta_{1}\cot\theta_{1}
-\sin\theta_{1})^{2}}{4(1+\cot^{2}\theta_{1})} .
\end{equation}
This can be compared to the failure probability when both qubits are
measured together, which corresponds to the case considered by Ivanovic,
Dieks and Peres
\begin{equation}
p_{fidp}=|\langle\Psi_{0}|\Psi_{1}\rangle |=\frac{1}{2}(\sin\theta_{1}
+\cos\theta_{1}) .
\end{equation}
These probabilities are plotted as a function of $\theta_{1}$in Fig.\ 1, 
and it can be seen that, as expected, $p_{f}\geq p_{fidp}$.  The 
probabilities are equal at some isolated points, but, in general, there is
a cost, which manifests itself as a higher failure probability, associated 
with determining the state by performing independent measurements on the 
two particles.  This example differs from our previous one in that here
there is a difference between $p_{f}$ and $p_{fidp}$, whereas there is none
when the two states we are trying to distinguish share the same Schmidt
basis.
\begin{figure}
\epsfig{file=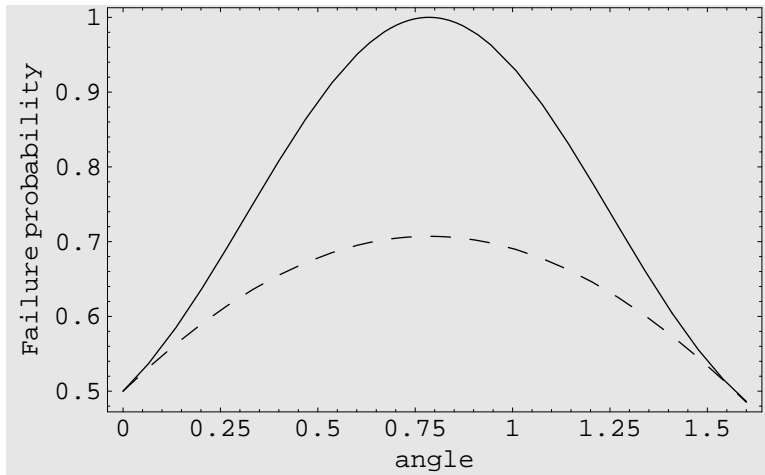}
\caption{Failure probabilities plotted as a function of the angle 
$\theta_{1}$.  The solid curve is $p_{f}$ and the dotted is $p_{fidp}$.
In this case the restriction on classical communication causes an 
increase in the failure probability.}
\end{figure}

\subsection{One failure state}
Let us now consider the case in which only one of the four measurement
alternatives corresponds to failure.  In particular, suppose that 
$\{ 0,0\}$ and $\{ 1,1\}$ correspond to $|\Psi_{0}\rangle$, $\{ 1,0\}$
corresponds to $|\Psi_{1}\rangle$, and $\{ 0,1\}$ corresponds to
failure.  We now have the conditions for our POVM operators
\begin{eqnarray}
A_{0}B_{0}|\Psi_{1}\rangle =0 &\hspace{1cm}& A_{1}B_{1}|\Psi_{1}\rangle
=0 \nonumber \\
A_{1}B_{0}|\Psi_{0}\rangle =0. & & 
\end{eqnarray}  
Using the same methods as before, we find that 
\begin{eqnarray}
A_{0}=|r_{A}\rangle\langle r_{A}| & A_{1}=|r_{A}^{\perp}\rangle\langle
r_{A}^{\perp}| \nonumber \\
B_{0}=|r_{B}\rangle\langle r_{B}| & B_{1}=|r_{B}^{\perp}\rangle\langle
r_{B}^{\perp}| .
\end{eqnarray}
Where we previously had two conditions on the orthonormal bases $\{ 
|r_{A}\rangle ,|r_{A}^{\perp}\rangle\}$ and $\{ |r_{B}\rangle ,|r_{B}^{\perp}
\rangle\}$, we now have three
\begin{eqnarray}
(\langle r_{A}|\otimes\langle r_{B}|)\Psi_{1}\rangle =0 & 
(\langle r_{A}^{\perp}|\otimes\langle r_{B}^{\perp}|)\Psi_{1}\rangle =0
\nonumber \\
(\langle r_{A}^{\perp}|\otimes\langle r_{B}|)\Psi_{0}\rangle =0 .
\end{eqnarray}

Let us now consider an example.  Let us assume that the states we are
trying to distinguish are given by Eq.\ (\ref{samebasis}), that is they 
have the same Schmidt basis.  Employing the same methods and notation
as before, we find first that $\theta_{1} =-\pi /4$, and that 
\begin{eqnarray}
z_{1}=-z_{4}^{\ast}=\sqrt{\tan\theta_{0}}
\nonumber \\
z_{2}=-z_{3}^{\ast}=\sqrt{\cot\theta_{0}}
\end{eqnarray}
The failure operator, $F$ is now
\begin{equation}
F=A_{0}^{\dagger}A_{0}B_{1}^{\dagger}B_{1}=|r_{A}\rangle\langle r_{A}|
\otimes |r_{B}^{\perp}\rangle\langle r_{B}^{\perp}| ,
\end{equation}
where
\begin{eqnarray}
|r_{A}\rangle = \left(\frac{1}{1+\tan\theta_{0}}\right)^{1/2}(|0\rangle
+\sqrt{\tan\theta_{0}}|1\rangle ) \nonumber \\
|r_{B}^{\perp}\rangle = \left(\frac{1}{1+\tan\theta_{0}}\right)^{1/2}
(|0\rangle-\sqrt{\tan\theta_{0}}|1\rangle ) .
\end{eqnarray}
If both states are equally probable, then the failure probability for this 
procedure is given by
\begin{eqnarray}
p_{f}& =& \frac{1}{2}(\langle\Psi_{0}|F|\Psi_{0}\rangle + \langle\Psi_{1}|
F|\Psi_{1}\rangle ) \nonumber \\
 & &=\frac{1}{2}(\cos\theta_{0}-\sin\theta_{0})^{2}+\frac{1}{4} .
\end{eqnarray}
This probability and $p_{fidp}$ are plotted as a function of as a function
of $\theta_{0}$ ($\theta_{1}$ has been set equal to $-\pi /4$) in Fig.\ 2.
\begin{figure}
\epsfig{file=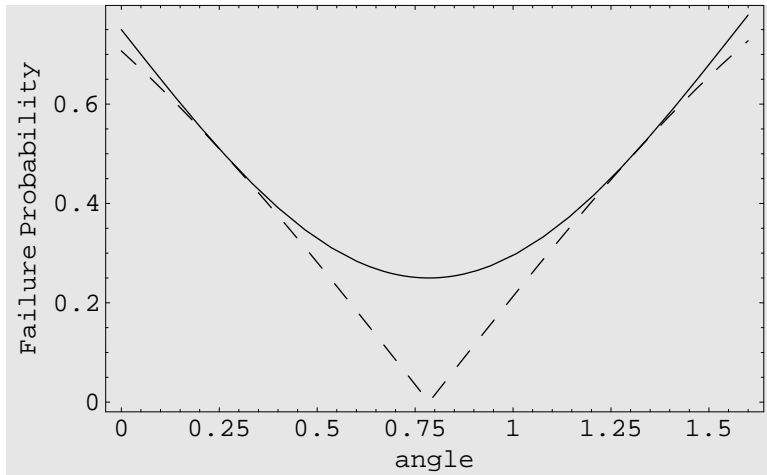}
\caption{Failure probabilities plotted as a function of the angle 
$\theta_{0}$ for the case of one failure state.  The solid curve is
$p_{f}$ and the dotted one is $p_{fidp}$.}
\end{figure}

\section{Secret sharing}
There have been a number of theoretical proposals for quantum secret sharing,
and one experimental demonstration.  The proposals fall into two categories.
In the first, quantum mechanics is used to securely distribute a classical
shared key.  One of these protocols is based on the use of GHZ states 
\cite{hillery} and another makes use of pairs of Bell states in different 
bases \cite{karlsson}.  An experiment based on the GHZ state protocol was 
carried out by  Tittel, Zbinden and Gisin \cite{gisin}.  The second category 
consists of protocols
in which the secret information that is split among several parties is 
quantum information \cite{gottesman}. The procedure we are considering
here is of the first type.

Let us suppose that a third party, Charlie, sends one of two states to
Alice and Bob, one qubit to Alice and one to Bob,
\begin{eqnarray}
|\Psi_{0}\rangle & = &\sin\theta |00\rangle + \cos\theta |11\rangle
\nonumber \\
|\Psi_{1}\rangle & = & \cos\theta |00\rangle +\sin\theta |11\rangle .
\end{eqnarray}
The procedure we are discussing here is based on the first example in
the preceding section.
Initially we shall suppose that Alice measures her state in the basis
given by
\begin{eqnarray}
|r_{A}\rangle & = & \frac{1}{(1+\cot\theta )^{1/2}}(|0\rangle +
\sqrt{\cot\theta}|1\rangle )\nonumber \\
|r_{A}^{\perp}\rangle & = & \frac{1}{(1+\tan\theta )^{1/2}}(|0\rangle
-\sqrt{\tan\theta}|1\rangle ) ,
\end{eqnarray}
and that Bob measures his particle in the basis
\begin{eqnarray}
|r_{B}\rangle & = & \frac{1}{(1+\cot\theta )^{1/2}}(|0\rangle -
\sqrt{\cot\theta}|1\rangle )\nonumber \\
|r_{B}^{\perp}\rangle & = & \frac{1}{(1+\tan\theta )^{1/2}}(|0\rangle
+\sqrt{\tan\theta}|1\rangle ).
\end{eqnarray}
By comparing their measurement results, Alice and Bob can determine
what state Charlie sent, or that the procedure has failed.  
Individually, however, they will not be able to make this determination.
Hence, Alice and Bob together will share a key with Charlie, individually
they will not.

Let us now examine the security of this scheme with regard to eavesdropping,
and we will quickly see that we have to modify the simple procedure
in the previous paragraph.  The reason is that an eavesdropper, Eve, has
a perfect cheating strategy.  Eve simply captures the particles, and
performs the same measurement on them that Alice and Bob would perform.
She then sends particles to Alice and Bob consistent with her measurement
results.  For example, if she finds $|r_{A}\rangle$ and $|r_{B}\rangle$,
she knows the state is $|\Psi_{0}\rangle$, and she sends a particle in
$|r_{A}\rangle$ to Alice and a particle in $|r_{B}\rangle$ to Bob. Using
this approach, Eve will know the key and Alice, Bob, and Charlie will not
be aware of her presence.  

This strategy of Eve's can be eliminated if
Alice and Bob sometimes measure in the $\{ 0,1\}$ basis.  Each of them
chooses randomly, with some predetermined probability, in which basis
to measure. When they compare their results, they look at the instances
in which they both measured in the $\{ 0,1\}$ basis, to see if their
results were ever different.  If they were, they can conclude that an
eavesdropper was present.  This defeats the attack proposed for Eve 
in the previous paragraph, because while the states $|\Psi_{0}\rangle$
and $|\Psi_{1}\rangle$ have no components along the vectors $|01\rangle$
and $|10\rangle$, states such as $|r_{A}\rangle |r_{B}\rangle$ do.
That means that in order to avoid detection, Eve must send states lying
in the subspace spanned by $|00\rangle$ and $|11\rangle$, which also
means that she will not be able to control the results that Alice
and Bob get. This leads to her detection.  When she measures the state
she receives from Charlie and fails, then she has to guess which state
to send on to Alice and Bob.  Sometimes she will guess incorrectly, and
if Alice, Bob and Charlie publicly compare some fraction of their data,
they will notice discrepecies, e.g.\  Charlie will have sent $|\Psi_{0}
\rangle$, but Alice and Bob will have detected $|\Psi_{1}\rangle$.  These
discrepencies would not exist if Eve were not present, and their
presence gives her away.

Next, let us see whether this procedure protects against 
cheating.  Suppose that Bob is able to capture both qubits 
sent by Charlie.  He first chooses a basis.  If it is $\{ 0,1\}$, he
sends a particle to Alice in one of these two states, and throws
out the two-qubit state from Charlie (because of his basis choice the results
from this state will not contribute to the key). When it comes time to 
compare results with Alice, if Alice measured the particle Bob sent in the
other basis, the results are thrown out, and if she also measured in the
$\{ 0,1\}$ basis, Bob simply announces the result corresponding to the 
particle he sent her.  If Bob chose to measure in the $\{ r_{A},
r_{A}^{\perp}\}$ and $\{ r_{B},r_{B}^{\perp}\}$ bases, then, if he
finds $|\Psi_{0}\rangle$ he send Alice $|r_{A}\rangle$, if $|\Psi_{1}
\rangle$, he sends $|r_{A}^{\perp}\rangle$, and if he fails he sends either 
$|r_{A}\rangle$ or $|r_{A}^{\perp}\rangle$.  If this is one of the results
that is publicly compared, then if Bob's measurement succeeded, he announces
the same state as the one he sent to Alice, and if it failed, the opposite
state.  Using this method, he knows the key bits, and Alice and Charlie
do not know that he knows.  

It is possible to fix this somewhat if instead of sending the particles
to Alice and Bob simultaneously, Charlie first sends one particle to
one party, who measures it and tells Charlie over a public channel that
he or she has received and measured the particle.  Charlie alternates
sending the first particle to Alice and Bob.  Now, supposing as before
that Bob is the cheater, let us see what happens when the particle is
sent to Alice first.  Bob grabs the particle that has been sent to Alice,
but then he must send her a substitute.  If he sends her a particle in
one of the states $|0\rangle$ or $|1\rangle$, there is no problem, but
he cannot do this all of the time, because then no key bits would be
generated.  If he sends her a particle in either $|r_{A}\rangle$ or
$|r_{A}^{\perp}\rangle$, he can run into a difficulty.  Suppose he
sent her $|r_{A}\rangle$, and when he receives the second particle from
Charlie, he finds that the state Charlie sent was $|\Psi_{1}\rangle$,
which should correspond to Alice measuring $|r_{A}^{\perp}\rangle$.
If he is to avoid creating a detectable error, he must claim, if this
is one of the bits which is publicly revealed, that he measured 
$|r_{B}^{\perp}\rangle$, which corresponds to failure to distinguish.
This, however, means that there will be more cases of failure to
distinguish than there should be, and Alice and Charlie would be
alerted to the fact that the security of the key is questionable.

Instead of sending Alice a single paricle in a specific state, Bob
can send Alice one of two particles in a singlet state.  This,
however, does not help him.  From the particle remaining in his
posession, he cannot determine which measurement Alice made, because
his particle could be in one of four possible states, and these 
cannot all be orthogonal.

In summary, the procedure outlined here provides protection against
eavesdropping, and some protection against cheating.  The presence of
an eavesdropper leads to errors (misidentification of states) while the
presence of a cheater leads to an increased failure rate.

\section{Conclusion}
We have examined the problem of distinguishing between two two-qubit 
states without error by using local measurements and either no or 
limited classical communication.  In the first case we found that only 
one of the two states can be identified, the other generates a
failure indication.  In the second case, for some pairs of states
it is possible to identify the states with the lowest possible
failure probability (the IDP limit), and for others the failure
probability with limited classical communication is higher than
the optimal value.  Finally, we proposed a secret sharing scheme
based on the procedure using limited classical communication.

Natural generalizations of this work are to higher dimensions, to
more than two states, and to states with more than two particles.  Many 
of our results rely explicitly on the fact that we are considering qubits, 
and the extension to qudits is not straightforward.  For example,
we found that with bipartite qubit states it is not possible to
distinguish two non-orthogonal states without using classical
communication.  We could tell if we had one of the two, but if
the other state was sent our procedure would always fail.  However, 
if we consider qutrits, whose basis states are $|0\rangle$, $|1\rangle$,
and $|2\rangle$, then the two bipartite states
\begin{eqnarray}
|\Psi_{0}\rangle & = & \frac{1}{\sqrt{2}}(|0\rangle |0\rangle +
|2\rangle |2\rangle ) \nonumber \\
|\Psi_{1}\rangle & = & \frac{1}{\sqrt{2}}(|1\rangle |1\rangle +
|2\rangle |2\rangle ) ,
\end{eqnarray}
which are not orthogonal, can be distinguished without classical
communication.  Measuring in the basis $\{ |0\rangle , |1\rangle ,
|2\rangle \}$, Alice and Bob will always obtain the same  result,
and if they obtain $|0\rangle$, they know that $|\Psi_{0}\rangle$
was sent, if they obtain $|1\rangle$, then $|\Psi_{1}\rangle$ was
sent, and if they obtain $|2\rangle$, then they have failed.  The 
extension to more than two states also introduces new elements.  For 
example, Ghosh, et al.\ have shown that it is not possible to 
deterministically distinguish either three or four orthogonal two-qubit 
states using only local operations and classical communication \cite{ghosh}. 
This suggests that there is much still to be learned about distinguishing
multipartite states using local operations and classical communication.

\section*{Acknowledgment}
This research was supported by the National Science Foundation under
grant number PHY-0139692 and by a PSC-CUNY grant.

\section*{Appendix}
We now want to show that if no classical communication is permitted, then
at most one state can be identified.  We begin by using the
conditions on the states and POVM operators to derive additional,
simpler ones.  For example, we have that
\begin{equation}
A_{0}B_{0}|\Psi_{1}\rangle = 0 \hspace{1cm}A_{0}B_{1}|\Psi_{1}\rangle =0 .
\end{equation}
Acting of the first of these with $B_{0}^{\dagger}$, the second with
$B_{1}^{\dagger}$, adding, and making use of Eq.\ (\ref{ident1}), we
find that
\begin{eqnarray}
0 & = & A_{0}(I_{B}-B_{f}^{\dagger}B_{f})|\Psi_{1}\rangle \nonumber \\
 & = & A_{0}|\Psi_{1}\rangle ,
\end{eqnarray}
where, in going from the first to the second line, we noted that 
$A_{0}B_{f}|\Psi_{1}\rangle =0$.  Similarly we find that
\begin{eqnarray}
B_{0}|\Psi_{1}\rangle = 0 & A_{1}|\Psi_{0}\rangle = 0 \nonumber \\
B_{1}|\Psi_{0}\rangle = 0 .
\end{eqnarray}

The next step is to express the states $|\Psi_{j}\rangle$, where $j=0,1$,
in terms of their Schmidt bases (see Section III)
\begin{eqnarray}
|\Psi_{0}\rangle &=& \sum_{j=0}^{1}\sqrt{\lambda_{0j}}|u_{Aj}\rangle
\otimes |u_{Bj}\rangle \nonumber \\
|\Psi_{1}\rangle &=& \sum_{j=0}^{1}\sqrt{\lambda_{1j}}|v_{Aj}\rangle
\otimes |v_{Bj}\rangle .
\end{eqnarray}
Application of the two conditions on $|\Psi_{1}\rangle$ in the previous
paragraph imply:
\begin{description}
\item[i.] If $\lambda_{10}\neq 0$ and $\lambda_{11}\neq 0$, then $A_{0}
|v_{Aj}\rangle=0$, for $j=0,1$, and this implies that $A_{0}=0$.  We 
also have that $B_{0}=0$.
\item[ii.] If one of the $\lambda_{1j}$'s is zero, and we can assume, 
without loss of generality, that $\lambda_{11}=0$, then we have that
$A_{0}|v_{A0}\rangle = B_{0}|v_{B0}\rangle =0$.
\end{description}
Similarly, the two conditions on $|\Psi_{0}\rangle$ give us:
\begin{description}
\item[iii.] If $\lambda_{00}\neq 0$ and $\lambda_{01}\neq 0$, then
$A_{1}=B_{1}=0$.
\item[iv.] If $\lambda_{01}=0$, then $A_{1}|u_{A0}\rangle=B_{1}
|u_{B0}\rangle =0$.
\end{description}

We now have a number of cases to examine.  If conditions $(i)$ and 
$(iii)$ are true, the only nonzero operators are the failure operators,
so that the procedure fails all the time.  If conditions $(ii)$ and
$(iv)$ are satisfied we have that the POVM operators $A_{j}$ and
$B_{j}$ must be of the form
\begin{eqnarray}
A_{0}=|\xi_{A}\rangle\langle v_{A1}| & B_{0}=|\xi_{B}\rangle\langle v_{B1}|
\nonumber \\
A_{1}=|\eta_{A}\rangle\langle u_{A1}| & B_{1}=|\eta_{B}\rangle\langle
u_{B1}| ,
\end{eqnarray}
where the vectors $|\xi_{A}\rangle$, $|\xi_{B}\rangle$, $|\eta_{A}\rangle$, 
and $|\eta_{B}\rangle$ are as yet undetermined.

We now examine the consequences of the conditions $A_{0}B_{f}|\Psi_{0}
\rangle =0$ and $A_{1}B_{f}|\Psi_{1}\rangle =0$, or
\begin{eqnarray}
A_{0}|u_{A0}\rangle \otimes B_{f}|u_{B0}\rangle =0 \nonumber \\
A_{1}|v_{A0}\rangle \otimes B_{f}|v_{B0}\rangle =0 .
\end{eqnarray}
The first of these equations implies that either $\langle v_{A1}|u_{A0}
\rangle =0$, which further implies that, up to a constant of modulus one,
$|v_{A0}\rangle = |u_{A0}\rangle$, or that $B_{f}|u_{B0}\rangle =0$.  If
the first alternative is true, then both $A_{0}$ and $A_{1}$ acting on
either vector $|\Psi_{j}\rangle$ gives zero, and the measurement always 
fails.  If this
alternative is to be avoided, then we must have $B_{f}|u_{B0}\rangle =0$.
However, the second equation tells us that, if the measurement does not
always fail, that $B_{f}|v_{B0}\rangle =0$.  These conditions imply that
(assuming that $|u_{B0}\rangle \neq |v_{B0}\rangle$; if this is not true 
the measurement always fails) $B_{f}=0$.  We then have that $I_{B}=
B_{0}^{\dagger}B_{0}+B_{1}^{\dagger}B_{1}$, which can only be true if
$|v_{B1}\rangle =|u_{B0}\rangle$ or $|v_{B0}\rangle =|u_{B1}\rangle$,
so that $\langle\Psi_{0}|\Psi_{1}\rangle =0$.  Sumarizing, we can
say that if $(ii)$ and $(iv)$ are satisfied, which implies that 
$|\Psi_{0}\rangle$ and $|\Psi_{1}\rangle$ are product states, then
either they are orthogonal, or the measurement always fails.

Finally, let us see what happens if $(i)$ and $(iv)$ are true (the
final alternative, $(ii)$ and $(iii)$ being true is equivalent).
This implies that $A_{0}=B_{0}=0$, so that $|\Psi_{0}\rangle$ is
never detected, and that $|\Psi_{0}\rangle$ is a product state.
Using techniques similar to those in the previous paragraphs, we
find that
\begin{eqnarray}
A_{1}=|\eta_{A}\rangle\langle u_{A1}| & B_{1}=|\eta_{B}\rangle
\langle u_{B1}| \nonumber \\
A_{f}=|\xi_{A}\rangle\langle u_{A0}| & B_{f}=|\xi_{B}\rangle
\langle u_{B0}| ,
\end{eqnarray}
where the vectors $|\xi_{A}\rangle$, $|\xi_{B}\rangle$, $|\eta_{A}\rangle$, 
and $|\eta_{B}\rangle$ are undetermined unit vectors.  The final conditions
are given by using the above expressions in the equations $A_{1}B_{f}
|\Psi_{1}\rangle =0$ and $A_{f}B_{1}|\Psi_{1}\rangle =0$ to give
\begin{eqnarray}
(\langle u_{A1}|\otimes \langle u_{B0}|)|\Psi_{1}\rangle &=& 0 \nonumber \\
(\langle u_{A0}|\otimes \langle u_{B1}|)|\Psi_{1}\rangle & = & 0 .
\end{eqnarray}
An example satisfying these conditions is given in Section II.

\bibliographystyle{unsrt}

\end{document}